\documentstyle[prl,aps,epsf,twocolumn,amssymb]{revtex}
\newcommand{\gsim}{\mbox{\raisebox{-1.ex}{$\stackrel
     {\textstyle>}{\textstyle\sim}$}}}
\newcommand{\lsim}{\mbox{\raisebox{-1.ex}{$\stackrel
     {\textstyle<}{\textstyle \sim}$}}}
\begin{document}

\draft

% UNCOMMENT FOR TWO-COLUMN MODE
\twocolumn[\hsize\textwidth\columnwidth\hsize\csname
@twocolumnfalse\endcsname

\title{Brane Quintessence}

\author{Kei-ichi Maeda}
\address{Department of Physics  and Advanced Research Institute for
 Science and
Engineering,\\  Waseda University, Shinjuku, Tokyo 169-8555,  Japan 
\\[-.7em]~}

%\date{\today}

\maketitle

%======================================%
%<<<<<<<<<<<<< ABSTRACT >>>>>>>>>>>>>>>% 
%======================================%

\begin{abstract}
We propose a new quintessence scenario in the brane cosmology, 
assuming that a quintessence field $Q$ is confined in our 3-dimensional 
brane world.
With a potential $V(Q)=
\mu^{\alpha+4} Q^{-\alpha}~(\alpha\geq 2)$, we find  that  
the density parameter of the scalar field  decreases
 as $\Omega_Q \sim a^{-4(\alpha-2)/(\alpha+2)}$ in  the epoch
of quadratic energy density dominance, if
$\alpha\leq 6$.  This attractor solution is followed by the usual 
tracking quintessence scenario
after a conventional Friedmann universe is recovered.
With an equipartition of initial energy density, we find a natural and 
successful  quintessence model for
$\alpha\gsim 4$.
\end{abstract}

\pacs{04.50.+h, 98.80.Cq~ \hfill 
%astro-ph/??
}
%\vskip 2pc
\vskip 1pc
]

\vskip.5cm
%\section{Introduction}
One of biggest mysteries in the Universe is a cosmological constant, if it
exists\cite{Weinberg}. Recent observation of Type Ia Super Novae  suggests 
that the expansion of the  Universe is accelerating now\cite{SNIa}. This
acceleration could be explained by a cosmological constant with
$\Omega_{\Lambda} \sim 0.7$. Although a cosmological constant seems to be 
preferred from a view point of the age of the Universe and a structure
formation in the Universe, one may wonder why the cosmological
constant is almost the same order of magnitude as the present mass density of
the Universe. It is  very difficult to explain such value from a view point
of particle physics, which usually predicts much larger value of a vacuum
energy. If the observation is confirmed, we will face on a serious 
problem in fundamental physics. One of the way out would be the so-called
``quintessence" \cite{Caldwell_Dave_Steinhardt}, in which a potential of a
scalar field plays as 
 a decaying cosmological
constant\cite{Dolgov_Fijii,Ratra_Peebles}.  In the quintessence scenario, a scalar
field with some specific potential shows an interesting behavior  called ``tracking"
(or ``scaling") in its evolution\cite{Zlatev_Wang_Steinhardt,Ratra_Peebles}.
The energy of a scalar field tracks the radiation energy (or matter energy) 
for rather long time and then  eventually becomes dominant  after matter
dominant era. In order for this scenario to work well, a scalar field will  
catch up neither radiation nor matter so early. 
This naively means that   the initial energy of a
scalar field was very small.
In the quintessence models, however, the final value of quintessence energy
density is insensitive to the initial conditions because of its attractor
property.  
For example, a potential of $V=\mu^{\alpha+4} Q^{-\alpha}$
will catch up matter density late in the evolution of the Universe
for a wide range of initial conditions,
if $\mu$ is suitably chosen. 
However, this suitable choice of $\mu$ may need a finetuning\cite{Weinberg2}.
Some modified models have been proposed to solve this
problem\cite{Armendariz-Picon_Chiba,Fujii_Dodelson}.  Here, in order to
resolve this mystery, we propose a scenario based on a brane universe.

Recently, a new type of world view has been proposed, which is called a
brane world based on a
superstring
 or M-theory\cite{Arkani-Hamed,Horava_Witten,Randall_Sundrum}.
Our 3-dimensional universe is described by  a brane in a
higher-dimension\cite{Rubakov_Shaposhinikov}, and usual matter field and force
except for gravity  are confined on the brane.  
Among them, 
Randall and Sundrum's second model\cite{Randall_Sundrum} 
gives  an
interesting picture of gravity, i.e. although  the extra-dimension is not
compact, four-dimensional Newtonian gravity is recovered  in
five-dimensional anti-de Sitter spacetime (AdS$_5$) in low energy limit.
Since gravity in those brane world could be  quite different from the
4-dimensional Einstein theory, many authors   discussed interesting difference
from a conventional cosmological
model\cite{Arkani-Hamed2,brane_cosmology,brane_cosmology2,brane_cosmology3}.    The
quadratic term of energy-momentum appears and may be important in the early stage of
the universe, and dark ``radiation", which is constrained by a successful
nucleosynthesis, may also exist\cite{brane_cosmology2,Shiromizu_Maeda_Sasaki}.
In particular, the former  will change the expansion law of the Universe in the very
early stage, then we may expect  some important difference
 in a ``quintessence" scenario.

Here, we will show that the quadratic term will indeed 
change drastically the evolution of a scalar field and  its density parameter  will
decrease in time until the quadratic term becomes unimportant. This provides us a
successful  and natural scenario for a conventional quintessence model.

%\section{Cosmological Solution}

In this letter, we shall analyze a cosmological solution based on the
Randall-Sundrum model, although the similar result would be obtained in other
brane world models. Assuming  flat Friedmann-Robertson-Walker spacetime in our
brane world, we find  the effective Friedmann  equation as
follows\cite{brane_cosmology2,Shiromizu_Maeda_Sasaki}:\\
\begin{eqnarray}
H^2  = {\kappa_4^2 \over 3} \rho + {\kappa_5^4 \over 36}
\rho^2 +{{\cal C}\over a^4}
\end{eqnarray}
where 
$\kappa_4^2 = 8\pi G_N$
and $\kappa_5^2$ are 4- and 5-dimensional
gravitational constants, respectively, $H=\dot{a}/a$ is the Hubble parameter, 
and ${\cal C} $ is
a constant, which denotes ``dark"
radiation\cite{brane_cosmology2}. The 4-dimensional Planck mass 
$m_4\,(=\kappa_4^{-1} = 2.4\times 10^{18}$GeV) and the
5-dimensional one $m_5\,(=\kappa_5^{-2/3})$ will be used in the following
discussion. Here  the 4-dimensional cosmological constant is set to zero.
We also assume that all matter field including a scalar field $Q$ 
 are confined
on the brane.
As for the potential,  we consider   one of typical quintessence-type
potential, i.e. $V(Q) = \mu^{\alpha+4} Q^{-\alpha} $\cite{Ratra_Peebles},
although the present mechanism may work for other potentials. 
It is worth noting that this potential may be naturally derived in some
supersymmetric QCD with a fermion condensation.
In that case, $\alpha$ is given by the numbers of colors and of
flavors\cite{Binetruy}.
Since the energy density decreases when the universe expands, the quadratic
term ($\rho^2$) dominates in the early stage of the universe. The conventional
Friedmann universe is recovered after when the quadratic and the linear
terms are equal at $\rho
=\rho_c = 12\kappa_4^2 /\kappa_5^4 = 12 m_5^6/m_4^2$.
 Since we know well about the behavior of the scalar field
in the linear-term dominant stage, which is the conventional cosmological
model,   we first study  the behavior of the scalar field in  the
quadratic-term dominant stage. 
%\subsection{$\rho^2$ dominant stage}
In order to study the dynamics of a scalar field, we discuss two cases
separately; the radiation dominant era and the scalar-field dominant era.

First we analyze the case with radiation dominance.
The Friedmann equation is approximated as
$
H ={\kappa_5^2 \over 6} \rho_{\rm r} \sim a^{-4},
$
where we assume that the Universe is expanding ($H>0$).
This gives the expansion law of the Universe  
as $a \propto t^{1/4}$. Then the equation of motion for the scalar field is 
\begin{eqnarray}
\ddot{Q} +{3\over 4t}\dot{Q} -\alpha \mu^{\alpha+4} Q^{-(\alpha+1)} =0 .
\end{eqnarray}
We find
an exact solution for $\alpha<6$, that is 
\begin{eqnarray}
Q  \propto  t^{2\over \alpha +2} ~~~~~{\rm and}~~~~~~
\rho_Q   \propto    t^{-{2\alpha\over \alpha +2}} \propto
a^{-{8\alpha\over \alpha +2}}.
\label{sol_scalar}
\end{eqnarray}
The density parameter of the scalar field, which we denote
as $\Omega_Q$,  is 
\begin{eqnarray}
\Omega_Q = {\rho_Q \over \rho_Q+\rho_{\rm r}+\rho_{\rm m}}\sim 
{\rho_Q
\over \rho_{\rm r}} \sim \beta \left({a\over a_s}\right)^{-{4(\alpha-2)\over
\alpha +2}},
\label{evol_Omega1}
\end{eqnarray}
where $\beta\,(\leq 1)$ is a constant and $a_s$ is an ``initial" scale factor.
If  $\alpha > 2$, $\Omega_Q $ {\it decreases} with time, just contrary to 
 a tracking solution. The scalar field energy decreases faster than the
radiation energy. This is a new interesting feature for a quintessence because
smallness of a quintessence-field energy when the Universe enters the
conventional stage could
be dynamically obtained. We will show that it is really the case. 
If  $\alpha = 2$, 
$\Omega_Q$ is constant until the linear  term becomes dominant. This is
the so-called ``scaling" solution. 

We  also find that the above solution (\ref{sol_scalar}) is an attractor, and
can show that  any solutions in the radiation dominant era will eventually converge to
this attractor solution\cite{Maeda_Yamamoto}.
The constant $\beta$ then denotes $\Omega_Q$ when the Universe reaches this solution at
$a=a_s$.

One may wonder if the  scalar field initially dominates the radiation. 
If
$\alpha <2$,  the potential term will overcome
the kinetic term, leading to an inflationary universe  as 
$a \propto \exp[H_0t^{(2-\alpha)/2}]$, where $H_0$ is a constant determined by
$\mu$ and $m_5$\cite{Maeda_Yamamoto}. 
For the case with $\alpha=2$, we find a
power-law solution\cite{Maeda_Yamamoto}, i.e.
$a  \propto   t^p$ with $p={1\over 6} [1+{1\over
8}\left({\mu/ m_5}\right)^6 ] $ and  $
Q =  2\sqrt{2}m_5^{3/2}t^{1/2} .$
If $p>1$ (i.e. $\mu \gsim 1.85\, m_5$), we have a power-law inflationary solution,
which is an attractor of the dynamical system.  While, if $p<1/4$ 
(i.e.
$\mu \lsim 1.26\, m_5$), this solution is no longer an attractor, leading to  the
radiation dominant era discussed above. 
If $\alpha>2$, we can  show that a kinetic term of the scalar field will
always become dominant even if we start with a potential
dominance\cite{Maeda_Yamamoto}. 
 Since $\rho_Q \propto a^{-6}$ in the case of kinetic-term dominance, we
 find that the radiation will overcome the scalar-field energy and the
Universe will evolve into the radiation dominant era  discussed above.
Since the solution (\ref{sol_scalar}) is a unique attractor in the
radiation dominant era, any solution in
$\rho^2$-dominant stage will approach to this attractor
solution, if $2<\alpha<6$. 
Then the evolution of $\Omega_Q$ is given by Eq. (\ref{evol_Omega1}). 

%\subsection{$\rho$ dominant stage}
As the universe expands and the energy density decreases below 
$\rho_{\rm c}$, we find the conventional Friedmann universe, in which 
many authors  studied  quintessence
models
\cite{Caldwell_Dave_Steinhardt,Zlatev_Wang_Steinhardt,Armendariz-Picon_Chiba,Fujii_Dodelson,quintessence}.
During the radiation or matter dominant era, when   the Universe evolves as $a
\propto t^{1/2}$ or $t^{2/3}$,
 the tracking attractor solution in the present model
was found, giving the evolution of $\Omega_Q$ as 
\begin{eqnarray}
\Omega_Q &\sim&  \rho_Q/ \rho_{\rm r} \propto a^{8\over( \alpha
+2)}~~~~{\rm (radiation ~dominant)}
\nonumber \\
\Omega_Q &\sim& {\rho_Q / \rho_{\rm m}}  \propto a^{6\over( \alpha
+2)}~~~~{\rm (matter ~dominant)}.
\label{evol_Omega2}
\end{eqnarray}
In both radiation dominant  and matter dominant cases, the energy density of
the scalar field  decreases slower than that of radiation or matter fluid, 
and will
eventually overcome those,  resulting in an accelerating Universe.

%\section{Natural Quintessence Model}

From  (\ref{evol_Omega1}) and (\ref{evol_Omega2}), the present value of $\Omega_Q$ is
estimated as
\begin{eqnarray}
\Omega_{Q,0} &=&\beta \left({a_c\over
a_s}\right)^{-{4(\alpha-2)\over
\alpha +2}} \left({a_{\rm eq}\over
a_c}\right)^{{8\over
\alpha +2}}\left({a_0\over
a_{\rm eq}}\right)^{{6\over
\alpha +2}}\nonumber \\
&\sim& \beta \left({T_c\over T_s}\right)^{{4(\alpha-2)\over
\alpha +2}}\left({T_{\rm eq}\over T_c}\right)^{-{8\over \alpha
+2}}\left({T_0\over T_{\rm eq}}\right)^{-{6\over \alpha
+2}},
\end{eqnarray}
where $a_c$, $a_{\rm eq}$, and $a_0$ are scale factors at the time when the
quadratic term of energy density drops just below the linear term, when
radiation   density becomes equal to matter density, and the present
time, and 
$T_c$, $T_{\rm eq}$, and $T_0$ are corresponding  temperatures of the Universe,
respectively.
In the present brane quintessence model, $\Omega_Q$ first decreases  during
the quadratic-energy dominant stage, and  the scalar field energy could be
very  small when the conventional cosmology is recovered. 
This may explain naturally why the scalar field energy was so small 
in the early stage of
the Universe in the conventional quintessence scenario.

Now we discuss about  constraints on parameters to find a successful
quintessence scenario.
Since we do not know about initial condition, we set our initial time
when the attractor solution is reached ($a=a_s$).
In this case, we have only one unknown  parameter $\beta$ (the value of
$\Omega_Q$ at $a=a_s$).
Although we do not know the value of $\beta$, it should be smaller than unity
because the attractor solution appears only in the radiation dominant era.
If an equipartition for the energy of each particle is assumed at $a_s$, we expect
$\beta \sim 1/g$, where $g$ is a degree of freedom of particles.
$\beta$ could be smaller because the kinetic energy of the scalar field might
be dominant before the attractor solution.

Setting $T_0 =2.73$ K and  $T_{\rm eq} \sim 10^4$ K, and assuming
$\Omega_{Q,0} \sim 0.7$ and equipartition  at
$a_s$ (i.e. $\beta=1/g=0.01$),
we find a relation between $T_s$ and $T_c$, which is shown in Fig. 1 by
a solid line for each $\alpha$.
%%%%%%%%%%%%%%%%%%%%%%%%%%%%%%%
\begin{figure}[htbp] 
\begin{center} 
 \leavevmode
 \epsfysize=4.5cm
 \epsfbox{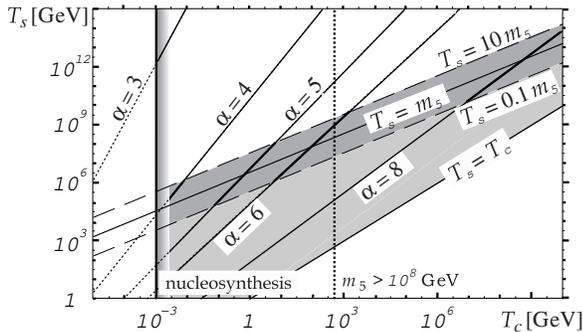}
\vskip 1em
\caption{A relation between $T_s$  and $T_c$ 
to find $\Omega_{Q,0} \sim 0.7$ under an initial equipartition condition for
$\alpha=3, 4, 5, 6$ and 8.  The nucleosynthesis constraint and  $m_5
= 10^8$ GeV are also shown by  solid and dotted vertical lines.
A successful scenario is obtained in the light and dark shaded regions. 
  The dark shaded
region ($0.1 \,m_5\leq T_s  \leq 10 \,m_5$) may be preferred for natural initial
condition.  }
 \label{fig1} 
\end{center} 
\end{figure}
%%%%%%%%%%%%%%%%%%%%%%%%%%%%%%%% 
One stringent constraint in any cosmological models is 
nucleosynthesis.
It must take place in the conventional radiation dominant era to explain the
present amount of light elements. Therefore,
$T_c$ must be higher than the temperature at nucleosynthesis, $T_{\rm NS}
\sim
$ 1 MeV. This constraint  implies 
$m_5 > 1.6 \times 10^4  \,(g/100)^{1/6}  (T_{\rm NS}/$1MeV$)^{2/3}$ GeV, which 
 is included in Fig. 1.
The r.h.s. of the vertical line is the allowed region.
 If the second  Randall-Sundrum 
model  turns out to  be a fundamental theory, in order to recover the Newtonian force
above 1mm scale,
the 5-dimensional Planck mass is constrained as $m_5
\geq 10^8$ GeV\cite{Randall_Sundrum}, 
which is satisfied in the r.h.s. region of the dotted vertical line in
Fig. 1.

Although we do not know the ``initial" temperature $T_s$, 
if the 5-dimensional
spacetime is fundamental  and  gravity is unified at the energy scale $m_5$, 
we expect  $T_s \lsim m_5$. Even if the 5-dimensional theory is effective,
 $T_s \lsim m_5$ may be required to justify the present 5-dimensional
analysis.  This condition with the equipartition at $a_s$ gives a constraint on 
the scale of the potential as $\mu \gsim (0.2-0.3) \,m_5$.
For reference, we have inserted  three lines of $T_s=0.1\, m_5$ (the lower
dotted line), $ m_5$ (solid line),
 and
$10 \, m_5$ (the upper dotted line) in Fig. 1.
If  $\alpha \gsim 4$, we find a successful quintessence scenario with natural
conditions.
On the other hand , if $\alpha<4$, either $\beta $ should be much smaller than
the value expected from equipartition, or $T_s \gg m_5$, then we may need a
finetuned or unnatural initial condition. 

In Fig. 2, we  show one example of the time variation of $\Omega_Q$ for the case of
$\alpha=5$ with 
$\beta=0.01$ and $T_s
= m_5$. 
For these parameters, we find that $T_c \sim $280 MeV
($\gg T_{\rm NS}
$), and
$T_s
(=
m_5) \sim 8.6 \times 10^6$ GeV. 

%%%%%%%%%%%%%%%%%%%%%%%%%%%%%%%
\begin{figure}[htbp] 
\begin{center} 
 \leavevmode
 \epsfysize=4cm
 \epsfbox{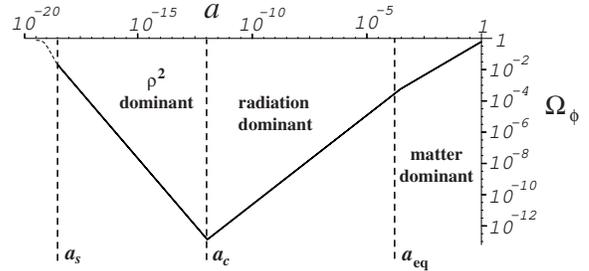}
\vskip 1em
\caption{ 
Time variation of $\Omega_Q$ 
 in terms of a scale factor $a$,
which present value is normalized to unity, for  the model with 
$V=\mu^{9}Q^{-5}$. We set 
$\beta =0.01$ and $T_s = m_5$. The energy of the scalar field drops faster than
that of radiation in the quadratic-energy dominant stage. After the conventional
Friedmann universe is recovered, the scalar field with very small energy density tracks
radiation and matter fluid.}
 \label{fig2} 
\end{center} 
\end{figure}
%%%%%%%%%%%%%%%%%%%%%%%%%%%%%%%%

If $\alpha\geq 6$, the above solution (\ref{sol_scalar})
is no longer an attractor.
We can show that the kinetic term  always dominates in the quadratic-term dominant
stage. Then $\Omega_Q $ drops as $a^{-2}$ until the conventional cosmology is
recovered at $a_c$. However, we know that the kinetic term drops much faster than the
potential term after $a=a_c$ and the tracking solution
eventually will be reached. Hence, we could approximate $\Omega_Q
$ at $a=a_c$  by the potential term, i.e $\Omega_{Q,c} \sim
\beta(a_c/a_s)^{4-\alpha}$. Using this estimation, 
we find the relation between $T_s$
and
$T_c$ to obtain
$\Omega_{Q,0} \sim 0.7$. Those results are also included in Fig. 1.

In the brane universe,  ``dark" radiation ${\cal C}\, (>0)$ may exist.
If this term dominates,  the present scenario may not work because it evolves as
$a^{-4}$ just as radiation, . The preliminary analysis shows that the quadratic
energy of the scalar field will drops as the same as the present model, but the
linear  term will increases because the expansion law of the Universe is $a\propto
t^{1/2}$. It will depend on the initial condition whether we find a successful model 
or
not.

In this letter, we have discussed a quintessence model in the context of a brane world
scenario. We show that the energy of a scalar field decreases  faster than
the radiation when the quadratic energy density is dominant. As a result,
when we recover a conventional cosmology, the density parameter of a scalar
field is enough small for a  scalar field to dominate just at
present epoch.  If our 3-dimensional brane universe starts at $T_s  \sim   m_5$,
$\alpha
\, \gsim
\, 4$ is required to find a natural quintessence scenario. 
We should note that the present model may not solve the coincidence problem, but
will give a natural explanation of smallness of a scalar field energy in the
early stage of the conventional cosmology.
It is also important to point out that we could predict when the Universe
is getting into an acceleration phase, if we know the values of fundamental
parameters such as
$m_5$ and $\mu$.

The present model  prefers rather large value of $\alpha$ ($\alpha \gsim 4$), but
recent observation may force to the constraint of $\alpha \lsim
2$\cite{Balbi_Baccigalupi_Matarrese_Perrotta_Vittorio}. If this is the case, we
 have to look for other type of quintessence potential\cite{quintessence}. For
example, an exponential potential could be another candidate.  The results
for such a study  will be published  elsewhere.

The essential point in the present scenario is that the dynamics of the scalar
field is completely modified in the quadratic energy dominant
stage\cite{brane_cosmology,brane_cosmology2,Maeda_Yamamoto}. Although we assume
the Randall-Sundrum model in the present analysis,  the quadratic energy term
will usually appear in any brane universe
models\cite{brane_cosmology,brane_cosmology2}, then we  expect the similar effect,
which may provide us a successful quintessence. 
Since the dilaton field will also
appear in a superstring or M-theory, it will also change the dynamics  of the
Universe\cite{brane_cosmology,brane_cosmology3}, which requires 
 further analysis. 

%\acknowledgments

KM would like to thank T. Harada,  S. Mizuno,  A. Starobinsky and
K. Yamamoto
 for useful comments and discussions. This work is supported partially
by the Grant-in-Aid for Scientific Research Fund of the Ministry of
Education, Science and Culture (Specially Promoted Research No.
08102010)  and by the Yamada foundation.

\end{document}